\documentclass[oribibl, final]{llncs}

\usepackage{graphicx}
\usepackage{makeidx}
\usepackage{amsmath}
\usepackage{verbatim}
\usepackage{blindtext}
\usepackage{pseudocode}
\usepackage{listings}
\usepackage{array}

\begin{document}

\title{Distributed Business Processes -- A Framework for Modeling and Execution}
\subtitle{A S-BPM Implementation based on Windows .NET Workflow Technology}

\author{Johannes Kotremba\and Stefan Ra\ss\and Robert Singer\thanks{Corresponding author}}
\institute{FH JOANNEUM -- University of Applied Sciences,\\
Alte Poststra\ss e 147, 8020 Graz, Austria\\
\email{johanneswalter.kotremba.aim11@fh-joanneum.at}\\
\email{stefan.rass.aim11@fh-joanneum.at}\\
\email{robert.singer@fh-joanneum.at}}

\maketitle

\begin{abstract}
Commercially available business process management systems (BPMS) still suffer to support organizations to enact their business processes in an effective and efficient way. Current BPMS, in general, are based on BPMN 2.0 and/or BPEL. It is well known, that these approaches have some restrictions according modeling and immediate transfer of the model into executable code. Recently, a method for modeling and execution of business processes, named subject-oriented business process management (S-BPM), gained attention. This methodology facilitates modeling of any business process using only five symbols and allows direct execution based on such models. Further on, this methodology has a strong theoretical and formal basis realizing distributed systems; any process is defined as a network of independent and distributed agents -- i.e. instances of subjects -- which coordinate work through the exchange of messages. In this work, we present a framework and a prototype based on off-the-shelf technologies as a possible realization of the S-BPM methodology. We can prove and demonstrate the principal architecture concept; these results should also stimulate a discussion about actual BPMS and its underlying concepts.
\keywords{Business Process, Distributed Software, BPM, BPMS, S-BPM, Workflow, WF, .Net, Agent, Communication, CCS, ASM}
\end{abstract}

\section{Methodology}

\subsection{Motivation}

Business process management (BPM) is still in the focus of business and computer sciences. This can be simply concluded from the number of conferences and publications regarding this topic. This is astonishing after approximately 30 years of practice and research, and seems to be driven mainly by technological interests generated by the motivation of how to support the "idea" of BPM with adequate technology.

The tradition of workflow -- as defined by the \emph{Workflow Management Coalition}\footnote{\url{www.wfmc.org}} -- research seems to be strongly influenced by methods, which mainly support "algorithmic" processes over "heuristic" processes. What we call algorithmic processes, are processes with a well pre-defined process logic. For example, BPMN is a well suited \emph{de facto} standard\footnote{\url{www.omg.org}} to define a semantic model of these type of processes; additionally, petri-nets (or methods derived from petri-nets) establish a broad set of methods to study such processes in depth. If everything was fine, we wouldn't need to talk about it. In real world enterprizes we can find a substantial amount of what we call "heuristic" processes. This type of process is human centric and heavily based on communication as the central aspect of coordination of work and mainly knowledge driven. These processes build process choreographies -- i.e. distributed communicating agents (multi-agent environment).

There seem to be some open questions in the domain of BPM, which need further research and answers. For example, Smith and Fingar~\cite{Smith:2003zs,Smith:2004fy,Smith.2007} proposed a paradigm change when they discussed the topic of $\pi$-calculus as foundation of business process (management). This has also been discussed, for example,  in~\cite{Puhlmann:2005lf,Puhlmann:2006mp}. Recently, also B\"orger~\cite{Borger:2011ib} discussed fundamental failings of traditional methodologies and Olbrich~\cite{Olbrich:2011zi} pointed out, that BPM "has failed to deliver". Again, we reassemble at the famous "business-IT divide". A process model is not BPM, and enterprizes are still in trouble to implement BPM and bring it to life.

There also seems to be a clear difference of the business process maturity level between large-scale enterprizes and small and medium enterprizes (SME); it is not easy to find any formal maturity level in SMEs as reported in~\cite{Feldbacher:2011kf}. One possible reason is, that there are still different views and a different understanding about BPM~\cite{BPTrends.2012}; another reason is, that the tools and methodologies are not well suited for many types of processes, situations and business process management systems (BPMS) to support the whole BPM-life-cycle (see for example~\cite{Weske.2012}). Especially, process execution seems to be away from maturity. It is an obvious matter of fact, that the import/export of BPMN 2.0 models between different tools is a cumbersome work, execution even more (see for example~\cite{Silver.2012a, Silver.2011}). The BPMN 2.0 standard document defines -- beside others -- a \emph{Common Executable} subclass. This subclass does not include the elements \emph{Pool}, \emph{Lane}, \emph{Message Flow} and \emph{Data Store} which are used for modeling collaborations; no wonder that this leads to a lack of support for execution.

These are the reasons why we want to discuss an alternative approach, called subject-oriented business process management (S-BPM). S-BPM offers a coherent approach from modeling to execution, which can be done by process owners without sophisticated IT knowledge. This includes an alternative way to model business processes, as discussed in the following section, and we present a solution based on off-the-shelf infrastructure to demonstrate elementary and enhanced functionality of such an enterprize BPMS, based on communicating agents.

\subsection{Subject-oriented Methodology}

\subsubsection{Foundation}

To overcome many problems of traditional BPM and BPMS, we propose an additional approach which is called subject-oriented business process management (S-BPM). S-BPM is included in the Gartner\footnote{\url{www.gartner.com}} Hype cycle for BPM since 2011 and categorized as technology trigger. A comprehensive discussion of S-BPM can be found in~\cite{Fleischmann:2012va}; a more hands-on approach, based on didactical concepts for teaching, can be found in~\cite{Fleischmann:2013}. Actually, one commercial implementation of the S-BPM concept is available from the company Metasonic\footnote{\url{www.metasonic.de}}; The Metasonic Process Suite is based on the well known IDE "Eclipse" and uses Java as a programming language; it creates Java source code according to the process model, which is then compiled and executed. We will present a more scalable realization concept which is quite different. Our solution is built on platform technology (see section~\ref{sec:WF}), which is based on precompiled classes and does not need runtime compilation.

\subsubsection{Formalization}

The core concept of S-BPM is based on the formal method of the \emph{Calculus of Communicating Systems} (CCS), as discussed in~\cite{Milner.1980}, and on concepts such as the \emph{Parallel Activity Specification Scheme} (PASS), firstly discussed in~\cite{Fleischmann.1994}. A formal proof can be found in~\cite{Borgert.2011}, which also discusses enhancements based on the $\pi$-calculus (an enhancement of CCS). S-BPM can also be expressed as \emph{Abstract State Machines} (ASM), as elaborated in~\cite{Fleischmann:2012va} by B\"orger. For an in-depth view on ASM see~\cite{Borger:2003}. Another formal representation can be done via the definition of a domain specific language, as discussed in~\cite{Hover.2013}.

\subsubsection{Execution}

S-BPM is based on strong formal building blocks, making it reasonable to translate S-BPM models into machine language without a lot of tricky work by software engineers. Beside the technical aspects, S-BPM offers an easy to understand methodology based on human communication to model business processes. In that way, we understand business processes as realization of structured communication between independent agents in a multi-agent system and a corresponding BPMS supports agents to execute the behavior models.

Based on the conceptual underlyings, we will furthermore present a general framework as a foundation for a S-BPM modeling and execution system; we will also present and discuss a prototype implementation as a basis for further work.

\subsubsection{Basic Building Blocks}

\begin{figure}
    \centering
    \includegraphics[scale=0.5]{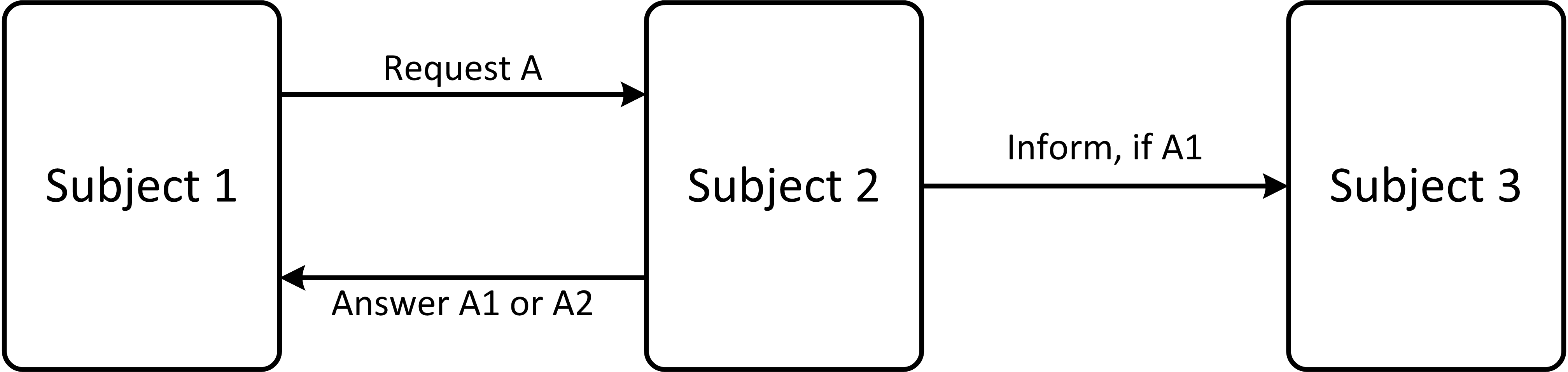}
    \caption{An example of a \emph{Subject Interaction Diagram}: it contains three subjects and all exchange messages. Subject 1 starts the process by sending a request to Subject 2; Subject 2 makes a decision, sends an answer back to Subject 1 and, in case of a positive answer, informs Subject 3 about the decision.}
    \label{fig:SID}
\end{figure}

In this chapter we give a short overview of the principal structure of a S-BPM model to have all necessary informations available for further discussions. Fig.~\ref{fig:SID} shows a S-BPM model, which contains the related subjects and the explicit communication relationships; this view is referred to as \emph{Subject Interaction Diagram} (SID) or, synonymously, as \emph{Communication Structure Diagram} (CSD). In general, message exchange means sending/receiving of \emph{Business Objects} (BO).

The possible sequences of a subject's actions in a process are termed subject behavior, i.e. the internal behavior of a subject. Related to other subjects, the internal behavior is not visible or known, as only the exchanged messages are relevant (black box or information hiding principle). This is an important part of the concept, as each subject has an autonomous behavior and therefore can be executed as agent (human or machine), reflecting the distributed nature of S-BPM. This is not a new principle, as the programming languages Occam-$\pi$ or Erlang and the transputer microprocessor architecture implement very similar concepts. Contrary to specific approaches we will present a solution build on off-the-shelf infrastructure, which has many advantages for practical use.

\begin{figure}
    \centering
    \includegraphics[scale=0.5]{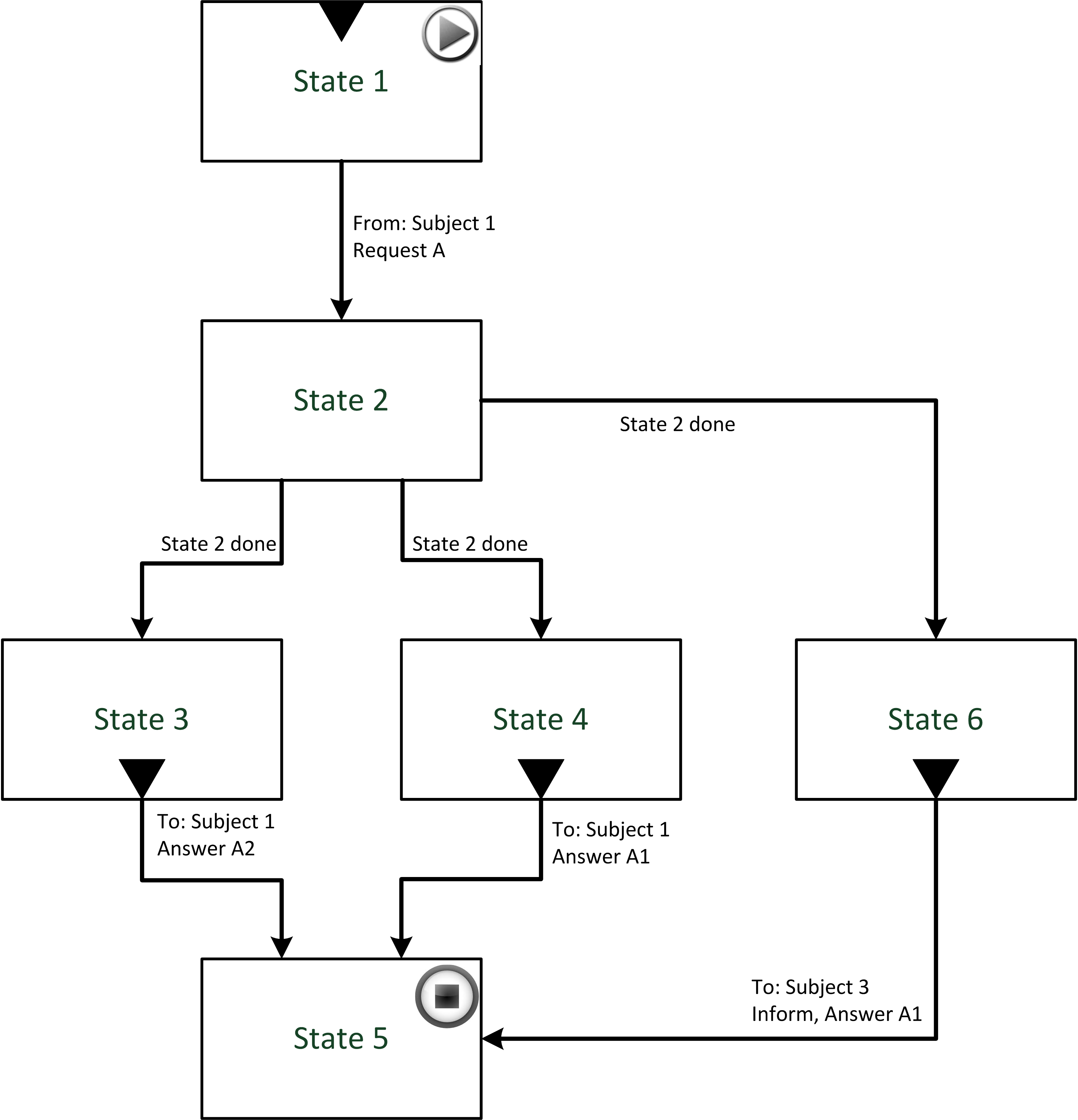}
    \caption{An example of a \emph{Subject Behavior Diagram}: triangles at the top symbolize \emph{Sending}, triangles at the bottom \emph{Receiving States}; other states are \emph{Function States} and states in general are marked as starting or ending state (by play and stop icon). State transitions are modeled as directed arcs.}
    \label{fig:SBD}
\end{figure}

To model the subject behavior we need states and state transitions which describe what actions a subject performs and how they are interdependent. To model any behavior we only need \emph{Send}, \emph{Receive} and \emph{Function states}; a \emph{function state} models any internal actions, which leads to another state of the subject. Any state has flags to indicate, if it is a starting or ending state. Fig.~\ref{fig:SBD} shows a possible \emph{Subject Behavior Diagram} (SBD) for Subject 2 in Fig.~\ref{fig:SID}.

\subsection{Workflow Technology}\label{sec:WF}

The main objective of the realization project was to develop a unified and scalable solution within a feasible timeframe, based on off-the-shelf infrastructure solutions. Additionally, we wanted to design a prototype which could be used as the base to develop a commercial solution, meaning it must include the full technology range typical for modern industry solutions. This decision was also based on industry input.

Therefore, we chose \emph{Microsoft Windows Server} as platform that provides an easy install of the Microsoft .NET framework, which includes the Microsoft \emph{Windows Workflow Foundation} (WF)~\cite{Chappell.2009} as a workflow engine, offering a bulk of needed functionality. Another advantage is, that it is also used by other Microsoft solutions, such as \emph{Microsoft Sharepoint} (since 2007) or \emph{Microsoft Dynamics CRM} (since 4.0). Technically speaking, \emph{WF} is a library within the .NET framework and can be used with programming languages like C\#; this offers a lot of possibilities for development and research.

A WF workflow provides functionality to maintain state, get input from and send output to the outside world, provides control flow, and executes code -- this is done by so called \emph{Activities}. The WF workflow class itself is also derived from the regular \emph{Activity} class.

\begin{figure}[t]
    \centering
    \includegraphics[scale=0.4]{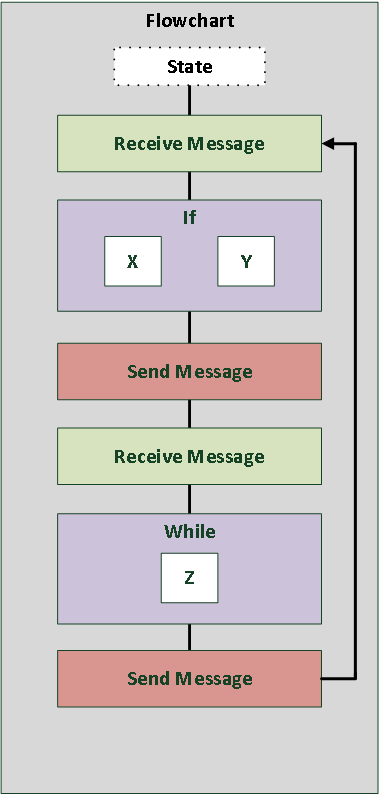}
    \caption{The structure of a WF workflow; all work is done by activities (modified from~\cite{Chappell.2009}). The \emph{Flowchart Activity} is enacted by the WF runtime engine and process flow can be routed back to previous \emph{Activities}.}
    \label{fig:WF_1}
\end{figure}

As Fig.~\ref{fig:WF_1} shows, every workflow has an outer activity that contains all of the others (e.g. a \emph{Sequence} or a \emph{Flowchart}). A \emph{Flowchart}, like other \emph{Activities} too, can contain variables that maintain its state and can contain other \emph{Activities}. Each \emph{Activity} in a WF workflow is actually a class. Execution of the workflow is performed by the WF runtime. The runtime does not know anything about the internal structure of an \emph{Activity}, but knows which \emph{Activity} to run next. It is important  to understand, that WF does not define any new language. To make life easier, WF includes a \emph{Base Activity Library} (BAL), which offers typical classes for workflows (as used in Fig~\ref{fig:WF_1}).

If not needed -- e.g. waiting for a message -- the state of a workflow can be persisted (into a persistence store) and stored safely until the continuation condition (e.g. an arriving message) is met. Based on this, a workflow might run on different threads, in different processes, and on different machines during its lifetime -- an important functionality for long-running processes. Any application build on WF technology is therefore scalable, since it is not confined to a single process on a single machine. Furthermore, another activity from the BAL offers parallel execution of child activities, another important functionality needed for a general workflow technology.

If we simply compare the representation of a SBD (see Fig.~\ref{fig:SBD}) with the representation of the WF workflow (see Fig.~\ref{fig:WF_1}), we can conclude, that it may be possible to map any SBD to a WF workflow. This will be our first topic to study and we will show that this can be done. For the S-BPM methodology to work in WF, custom activities are needed to perform the functionality of the S-BPM states: a custom \emph{Function}, \emph{Receive} and \emph{Send Activity}.

\section{Communicating Workflows}

As we do not only have one subject and therefore one WF workflow, we also have to develop an underlying concept to manage workflow interaction. This will be the first step in the development of our S-BPM architecture. After that, we have to discuss how to design or import a S-BPM process model.

\subsection{S-BPM Architecture Concept}

\begin{figure}
    \centering
    \includegraphics[scale=.33]{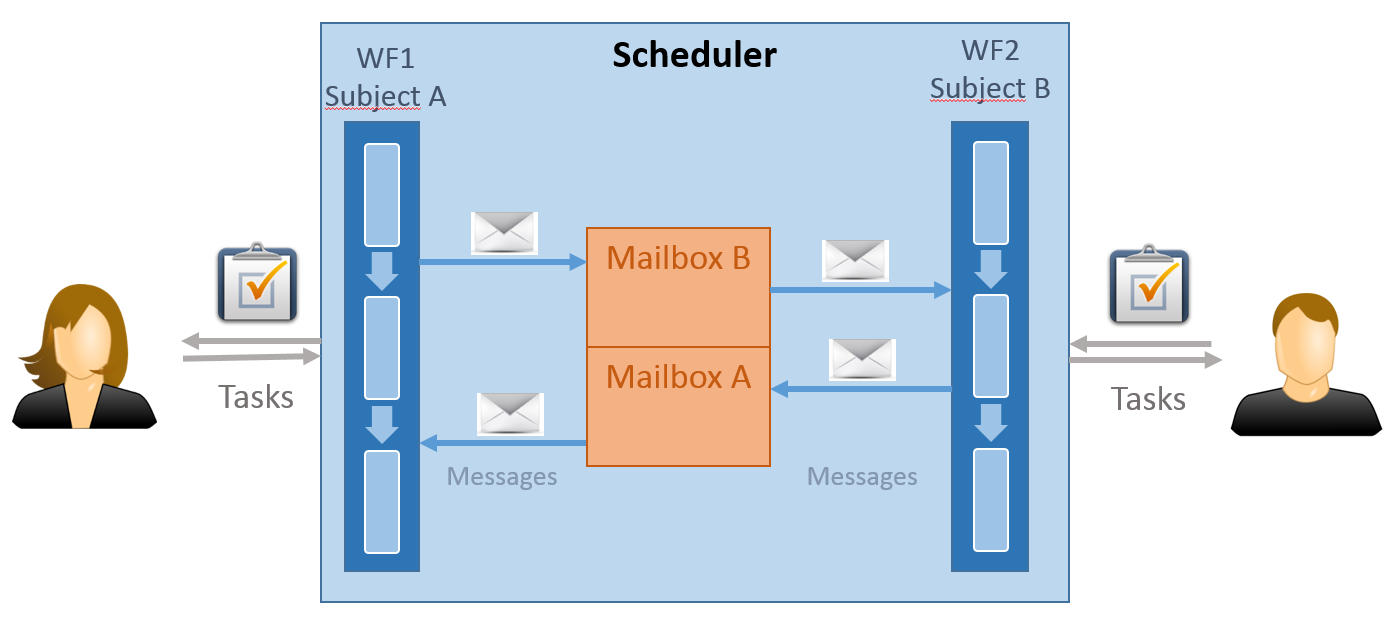}
    \caption{The Figure shows the execution of a process with two subject instances, i.e. agents (Subject A and Subject B).  The behavior of each subject is defined by a sequence of custom activities defined by a WF workflow (WF1 and WF2). The workflow activities can basically perform three different actions: send a message, do something, and receive a message. Each subject has its own message pool. A workflow communicates with clients in the form of \emph{Tasks}.}
    \label{fig:ArchConcept}
\end{figure}

The basic component of our architecture model is an application titled \emph{Scheduler} (see Fig.~\ref{fig:ArchConcept}). The \emph{Scheduler} represents the server-side execution environment for processes, while all necessary interactions with users are performed on the client side.

\subsubsection{The Message Pool}

The message pool concept is a central mechanism of \mbox{S-BPM}; in S-BPM all subjects have their own input message pool and message exchange between subjects can be synchronous or asynchronous. We need both types, as subjects are instantiated as agents and an agent can be a human or a machine, respectively a service. Further on, a subject has full access to all messages in its input pool and it can remove any of these messages for processing. This is a fundamental functionality for real world business processes, reflecting the fact, that a knowledge worker decides which process to continue next (in general it allows settings priorities). More details about the message concept can be found in~\cite{Fleischmann:2012va}.

So any agent can send messages to the message pool of another agent and take out messages from its own message pool. Workflow activities can require user interaction. In our implementation concept the user-interaction is performed client-side. Therefore the \emph{Scheduler} generates a \emph{Task} for the responsible agent and also includes corresponding data fields (read and/or write); in case of a human agent (user-task) this will typically lead to a form to be completed and returned to the \emph{Scheduler}. A possible realization will be discussed in section~\ref{sec:Prototype}

\subsubsection{The (Task) Scheduler}

The \emph{Scheduler} is the core component of our S-BPM architecture and has the following functionality:

First of all, the \emph{Scheduler} is acting as a host environment for all WF workflows. Each instance of a S-BPM process consists of several communicating subject instances (agents). The \emph{Scheduler} manages loading, instantiating, termination, unloading, and the storage of workflows, including the synchronous or asynchronous execution of workflows.

Furthermore, the \emph{Scheduler} manages the message exchange between the subject instances (agents). Messages can be exchanged by the use of specifically designed activities from within the WF workflows. The \emph{Scheduler} takes care that messages are delivered to the dedicated recipients.

Based on our server-client concept, activities within a WF workflow can require user interaction. For that reason, our specifically designed activities communicate with their clients in the form of \emph{Tasks}.

Summarized, the \emph{Scheduler} acts as WF workflow host, enables the communication between agents, delivers \emph{Tasks} from the WF activities to the clients and receives completed \emph{Tasks} from the clients to deliver them to their corresponding WF activity.

\subsection{S-BPM as Windows Workflow Model}

WF workflows can be serialized as data structure or files using the \emph{Extensible Markup Language} (XAML). XAML is a Microsoft proprietary declarative language used in technologies like the \emph{Windows Presentation Foundation} (WPF), \emph{Silverlight} or the \emph{Windows Workflow Foundation} (WF). XAML is a XML based language and in the background it is executed as program code, e.g. C\#.

The mapping of a SBD onto a WF workflows can be done in the following way: there are four elements in S-BPM which need equivalents in WF workflows: subjects, states (send, receive, function), transitions and parameters (local, global). The WF equivalent for a subject in general is a WF \emph{Flowchart Activity}. In that case we cannot explicitly model transitions, so they are merged together with states into activities. Each S-BPM state and its following transition/s are a custom WF \emph{Activity}, as explained in the next section. Parameters in S-BPM are converted to variables in WF, which provide the same functionality. S-BPM parameters assigned to S-BPM states become WF variables assigned to WF activities.

\subsubsection{S-BPM WF Activities}

As we need WF \emph{Activities} with specific behavior, we only need to "enhance" the standard \emph{Activity} class (we use C\#) with additional functionality. An activity performing the actions of a S-BPM \emph{Function State} in WF needs to have the following specific properties: a list of parameters which are read only, a list of parameters which are editable, a list of transitions which contains all available transitions to other states, and a boolean indicating if the state is an end state. It is important to understand, that any \emph{Function State} can include so called \emph{Refinements}, that is any additional functionality, for example calling any web service.

An activity performing the actions of a S-BPM \emph{Send State} in WF needs to have the following specific properties: a list of parameters which are read only, a list of parameters which are editable, a string with the name of the subject to which the message is sent to, a string with the message type, a string containing all parameters which are sent with the message, and a boolean indicating, if the state is an end state. In case of a \emph{Send State Activity}, the transition property is replaced by the \texttt{toSubject} property, which, in combination with the \texttt{MessageType} property, performs the task of a transition.

\begin{figure}
    \centering
    \includegraphics[scale=0.8]{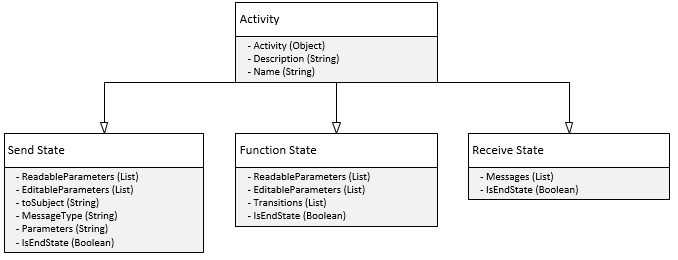}
    \caption{This Figure shows an UML class diagram of the custom WF \emph{Activities} derived from the main \emph{Activity} class.}
    \label{fig:cd_1}
\end{figure}

The \emph{Receive State Activity}  is the simplest of the triplet and has the following properties: a list of message types which can be received and a boolean indicating, if the state is an end state. In our implementation, a \emph{Receive State} can only have one transition, therefore there is no need for such a property. Table \ref{tab:sbpmwf} shows a summary of the mapping of S-BPM elements onto a WF workflow. Fig.~\ref{fig:cd_1} shows an overview of the created classes.

\begin{table}
\begin{tabular}{|>{\raggedright}p{4cm}|l|}
	\hline
	\textbf{S-BPM Element} & \textbf{WF Equivalent}\tabularnewline
	\hline
	\hline
	\emph{Subject} & \emph{WF Flowchart Activity}\tabularnewline
	\hline
	State and Transition & embedded in \emph{Function}, \emph{Send}, and \emph{Receive State Activity}\tabularnewline
	\hline
	\emph{Function State} & \emph{Function State Activity} (derived from Activity)\tabularnewline
	\hline
	\emph{Send State} & \emph{Send State Activity} \tabularnewline
	\hline
	\emph{Receive State} & \emph{Receive State Activity}\tabularnewline
	\hline
	\emph{Parameters} & \emph{Variables}\tabularnewline
	\hline
\end{tabular}
\smallskip\caption{This table shows a summary of the S-BPM elements and their WF equivalents.}\label{tab:sbpmwf}
\end{table}


\subsection{Process Repository}

All information about processes and their execution has to be stored in a proper way. Therefore, all defined processes as well as their running instances are stored within a central process repository on the server side. Additionally, we have to consider a mapping between organizational roles and subjects, i.e. a specific role (agent) is mapped onto a specific instance of a subject; roles are typically defined in the active directory structure of the IT infrastructure. Normally, one user can be assigned to several subjects and a subject can be assigned to several users.

\begin{figure}
	\centering
  	\includegraphics[scale=0.4]{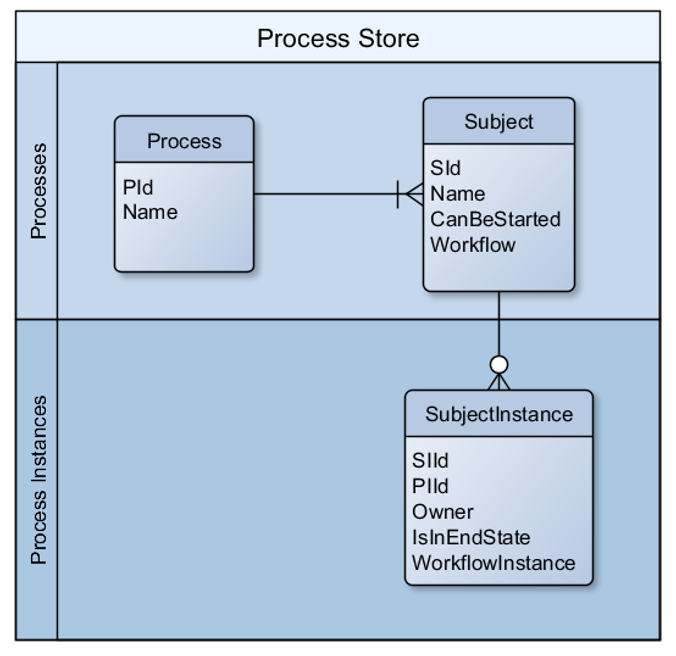}
	\caption{Entity Relationship Diagram (ERD) of the process repository data; relations between process, subjects, and subject instances.}
	\label{fig:processstore}
\end{figure}

Fig.~\ref{fig:processstore} shows the data relation between processes and their instances. In a very basic form each process has an identifier (the process identifier \texttt{PId}) and a name (\texttt{Name}). Each process contains at least one subject. Subjects are identified by their own identifier, the subject identifier (\texttt{SId}). Furthermore, each subject has several properties: \texttt{Name}, \texttt{CanBeStarted} and \texttt{Workflow}. The entire behavior of the subjects is stored as a XAML workflow within the property \texttt{Workflow}.

When a process is started, a new instance is created. All the required information is stored within the \texttt{SubjectInstance} entity. Each instance of a process needs a unique identifier, the process instance identifier (\texttt{PIId}). As well as each process instance, also each instance of a subject has a unique identifier (\texttt{SIId}). An instance of a subject is always connected to a user. Therefore the users own identifier or username is stored within the property \texttt{Owner}. The property \texttt{IsInEndState} describes whether a subject instance is currently within end state or not.

\subsection{Process Initialization}

To determine which user is allowed to start which processes, the following query (see listing~\ref{lst:listing1}) has to be resolved within the repository: find all subjects which are related to the user, where the subject's property \texttt{CanBeStarted} equals \texttt{true}.

\lstset{frame=lines, basicstyle=\small}

\lstset{language=SQL}
\begin{lstlisting}[caption={Repository query to determine, which processes can be started by a given user (\texttt{username}).},label={lst:listing1}]
/*Query 1*/
select s.SId, p.Name
	from Subjects as s
	inner join Users as u on u.Subjectname = s.Name
	inner join Processes as p on p.PId = s.PId
where s.CanBeStarted = true and u.Username = [username]
\end{lstlisting}

If a user wants to start a new process (see Listing~\ref{lst:listing2}), the \emph{Scheduler} needs to know the identifier of the subject starting the process and the username. In a first step the \emph{Scheduler} finds the corresponding subject for the corresponding identifier. Then a new instance of the subject is created based on this subject. This new instance gets a new unique identifier for the subject instance (\texttt{SIId}) and the process instance (\texttt{PIId}). Finally, a new instance of the subject's workflow is started.

\lstset{language=JAVA}
\begin{lstlisting}[caption={Pseudo code snippet to demonstrate process initialization based on SId and Username.},label={lst:listing2}]
StartNewProcessInstance(sId, username) {
  i = new SubjectInstance(sId, username);
  i.WorklfowInstance.run();
}
\end{lstlisting}

\subsection{Workflow Persistence and \emph{Tasks}}

Activities within a workflow can ask the user for interaction. As already mentioned, this interaction is handled by \emph{Tasks}. An activity creates a \emph{Task} and submits it to the user. After the \emph{Task} is answered by the user, the activity has all required information to continue its execution. There is a delay between creating a \emph{Task} and receiving the corresponding \emph{Task} answer. As the idle time can take from several minutes to even days, it is not needed that the entire workflow instance stays in the working memory of the server. For this purpose, \emph{Windows Workflow Foundation} includes a feature called workflow persistence. The feature makes a durable capture of a workflow instance's state, which can be used to resume the entire workflow later.

\begin{figure}
	\centering
  	\includegraphics[scale=0.4]{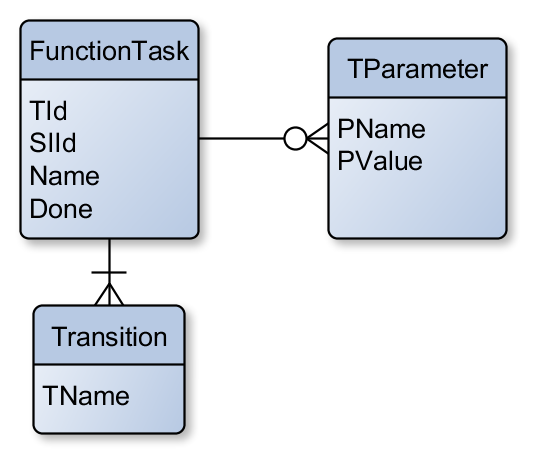}
	\caption{ERD of \emph{Function Task} data; relations between \emph{Function Task}, transitions, and parameters.}
	\label{fig:functiontask}
\end{figure}

For each type of a S-BPM state there exists an own \emph{Task} type, as described before. For instance, Fig.~\ref{fig:functiontask} shows the properties of a \emph{Function Task}. Each \emph{Task} has a unique identifier (\texttt{TId}) as well as a name. The name is typically the name of the activity which generated the \emph{Task}. Furthermore each \emph{Task} -- whether \emph{Function}, \emph{Send} or \emph{Receive Task} -- has the properties \texttt{SIId} and \texttt{Done}. The \texttt{SIId} defines to which subject instance the \emph{Task} belongs and the property \texttt{Done} marks whether a \emph{Task} has already been answered by the user or not. The \emph{Function Task} also stores all possible state transitions and the values of connected variables. The user may change values of some variables and one of the possible transitions, as defined in the underlying process model, has to be selected.

When the execution of a S-BPM state (which is a \emph{WF Activity}) starts, a new \emph{Task} is created. The new \emph{Task} is handed over to the \emph{Scheduler} which stores the \emph{Task} into the  repository. As soon as the \emph{Task} is stored into the repository, the \emph{Scheduler} starts persisting the workflow. During this step the current state of the workflow instance is captured and stored within the repository (property \texttt{WorkflowInstance}). Afterwards, the entire workflow instance is dropped from the systems working memory (sleeping).

To receive a list of open \emph{Tasks} for a specific user, it is necessary to find all open \emph{Tasks}, that are related to the user's subject instances (meaning all subject instances where the user is the owner). Open \emph{Function} and \emph{Send Tasks} will always be included in the \emph{Task} list (see listing~\ref{lst:listing3.2} and listing ~\ref{lst:listing3.3}), while \emph{Receive Tasks} (see listing~\ref{lst:listing3.4}) are only included, when additionally at least one message for that \emph{Task} exists.

\lstset{language=SQL}
\begin{lstlisting}[caption={Repository (pseudo) query to determine open \emph{Function Tasks} for a specific user.},label={lst:listing3.2}]
/*Query 2*/
select *
 	from FunctionTasks t
	inner join SubjectInstances i on t.SIId = i.SIId
	inner join Subjects s on i.SId = s.SId
where t.Done = false
	and (i.Owner = [Username]
	or (i.Owner = null and s.Name = ANY(
		select u.Subjectname from Users u
		where u.Username = [Username])))
\end{lstlisting}

\begin{lstlisting}[caption={Repository (pseudo) query to determine open \emph{Send Tasks} for a specific user.},label={lst:listing3.3}]
/*Query 3*/
select *
 	from SendTasks t
	inner join SubjectInstances i on t.SIId = i.SIId
	inner join Subjects s on i.SId = s.SId
where t.Done = false
	and (i.Owner = [Username]
	or (i.Owner = null and s.Name = ANY(
		select u.Subjectname from Users u
		where u.Username = [Username])))
\end{lstlisting}

\begin{lstlisting}[caption={Repository (pseudo) query to determine open \emph{Receive Tasks} for a specific user.},label={lst:listing3.4}]
/*Query 4*/
select *
 	from ReceiveTasks t
	inner join SubjectInstances i on t.SIId = i.SIId
	inner join Messages m on t.SIId = m.ToSIId
	inner join Subjects s on i.SId = s.SId
where t.Done = false and m.Received = false
		and (i.Owner = [Username] or
		(i.Owner = null and s.Name = ANY(
			select u.Subjectname from Users u
			where u.Username = [Username])))
		and m.MType = ANY(t.MTypes)
\end{lstlisting}

When a \emph{Task} is answered, the \emph{Scheduler} has to make sure, that the corresponding subject instance receives the answered \emph{Task} from the user. Therefore, the \emph{Scheduler} has to find the corresponding subject instance by its identifier, the \texttt{SIId}. As next step, the corresponding \emph{Flowchart Workflow} has to be resumed; then the answered \emph{Task} is forwarded to the workflow instance and the workflow is able to continue execution.

\subsection{Message Exchange}

\begin{figure}
	\centering
  	\includegraphics[width=220px]{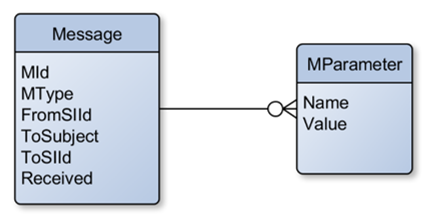}
	\caption{ERD of \emph{Message} data; relations between \emph{Message} and parameters.}
	\label{fig:message}
\end{figure}

Messages are used for the communication between agents. To enable an asynchronous communication between agents (i.e. subject instances), all messages are stored within the repository. Fig.~\ref{fig:message} shows the data definition of a message. Each message has a unique identifier (\texttt{MId}), a type (\texttt{MType}) and information about the sender and recipient (\texttt{FromSIId}, \texttt{ToSubject}, \texttt{ToSIId}). The property \texttt{Received} is used to mark whether a message is received by the recipient or not. Furthermore, messages can contain one or more parameters, which are defined by \texttt{Name} and \texttt{Value} pairs; as mentioned before, in general the concept assumes exchange of business objects.

\subsubsection{Sending a message}

Within a workflow, the \emph{Send Activity} is responsible for the creation of a new message object. A message has a message type (\texttt{MType}) and optionally one or more parameters. Moreover, the message also needs a recipient. Basically, the recipient is defined by its subject and username. Additionally, it is possible to define the recipient only by its subject. In this case each user, who represents this subject, is a possible recipient for this message.

Listing ~\ref{lst:listing4} shows the actions needed for sending and delivering messages: After the creation, the message is handed over to the \emph{Scheduler}. In a first step the \emph{Scheduler} stores the message into the recipient's message pool. In the next step the \emph{Scheduler} determines the process \texttt{PId} to which the sender belongs (see listing ~\ref{lst:listing5.5} ). Based on the \texttt{ToSubject} property the \emph{Scheduler} selects the corresponding subject \texttt{SId} within the process definition (see listing ~\ref{lst:listing5.6}). Now the \emph{Scheduler} has to check whether an instance of the receiving subject with the same \texttt{PIId} as the sender already exists or not (see listing ~\ref{lst:listing5.7}). In the case that no corresponding subject instance exists, the \emph{Scheduler} has to create a new one and starts its execution. The message property \texttt{ToSIId} will be updated with the corresponding \texttt{SIId} of this new subject instance. In case that an instance already exists, the message property \texttt{ToSIId} will be updated with the corresponding \texttt{SIId} of this subject instance.

\lstset{language=JAVA}
\begin{lstlisting}[caption={Pseudo code to demonstrate the send message operation.},label={lst:listing4}]
SendMessage(message) {
  processInfo = Query5(message.FromSIId);
  ToSId = Query6(processInfo.PId, message.ToSubject);
  ToSIId = Query7(ToSId, processInfo.PIId);
  if(ToSIId != null){
    message.ToSIId = ToSIId;
    repository.add(message);
  }
  else{
    i = new SubjectInstance(ToSId, processInfo.PIId);
    message.ToSIId = i.SIId;
    repository.add(message);
    i.WorkflowInstance.run();
  }
}
\end{lstlisting}

\lstset{language=SQL}
\begin{lstlisting}[caption={Repository query to determine the PId and PIId of a subject instance [FromSIId].},label={lst:listing5.5}]
/*Query 5*/
select s.PId, i.PIId
	from Subjects s
	inner join SubjectInstances i on i.SId = s.SId
where i.SIId = [FromSIId]
\end{lstlisting}

\begin{lstlisting}[caption={Repository query to determine the SId by its name and process identifier.},label={lst:listing5.6}]
/*Query 6*/
select SId
	from Subjects s
where s.PId = [PId]
	and s.Name = [ToSubject]
\end{lstlisting}

\begin{lstlisting}[caption={Repository query to determine the SIID for a given subject within a process instance.},label={lst:listing5.7}]
/*Query 7*/
select SIId
	from SubjectInstances i
where i.SId = [SId] and i.PIId = [PIId]
\end{lstlisting}

\subsubsection{Receiving a message}

All messages are stored within the message pool. To receive a message within a receive activity, the following steps need to be executed. First off all, it is necessary to find all messages which can be received by the current receive activity. These are all messages where the \texttt{ToSIId} equals to the current \texttt{SIId} of the subject instance and the message type (\texttt{MType}) is one of the types, which can be received by the current receive activity (see listing ~\ref{lst:listing6}). In case that the query finds a proper message, the message can be received by the receive activity. If there is more than one proper message, the users have to select the message they want to receive.

\lstset{language=SQL}
\begin{lstlisting}[caption={Repository query for receiving messages.},label={lst:listing6}]
/*Query 8*/
select MId
	from Messages m
where m.Received = false
	and m.MType = ANY([Message Types])
	and m.toSIId = [SIId]
\end{lstlisting}

After receiving the message, the message property \texttt{Received} has to be set to \texttt{true}. The subject instance will not continue the execution of its corresponding workflow until a proper message is received.

\subsection{Process Termination}

In S-BPM each state can be an end state. A process is ended, when each subject within the process is in an end state. Therefore the \emph{Scheduler} has to monitor all running process instances and terminate them if necessary.

While a subject instance is within an activity which represents an end state, its property \texttt{IsInEndState} is set to true. Each time a subject instance enters an end state, the \emph{Scheduler} executes the steps shown in listing ~\ref{lst:listing7}: First, the \texttt{IsInEndState} property of the subjects instance is set to \texttt{true}. Next, the \emph{Scheduler} checks whether all subject instances, which belong to the same process instance, have the property \texttt{IsInEndState} set to \texttt{true} or not (see listing ~\ref{lst:listing8.9}). In case all subject instances of the process instance are within an end state, all affected subject instances and the entire process instance will be terminated.

\lstset{language=JAVA}
\begin{lstlisting}[caption={Pseudo code to demonstrate the determination of the process end.},label={lst:listing7}]
SubjectInstanceEntersEndState(currentSubjectInstance) {
  currentSubjectInstance.IsInEndState = true;
  if(Query9(currentSubjectInstance.PIId) == 0) {
    foreach(i in Query10(currentSubjectInstance.PIId)) {
      i.WorkflowInstance.terminate();
    }
  }
}
\end{lstlisting}

\lstset{language=SQL}
\begin{lstlisting}[caption={Query supporting the determination of a process end.},label={lst:listing8.9}]
/*Query 9*/
select count(*)
	from SubjectInstances i
where i.IsInEndState = false
	and i.PIId = [PIId]
\end{lstlisting}

\begin{lstlisting}[caption={Query to determine all subject instances within a process instance.},label={lst:listing8.10}]
/*Query 10*/
select *
	from SubjectInstances i
where i.PIId = [PIId]
\end{lstlisting}

\section{S-BPM Prototype}\label{sec:Prototype}

\subsection{Architecture}

\begin{figure}
	\centering
  	\includegraphics[width=320px]{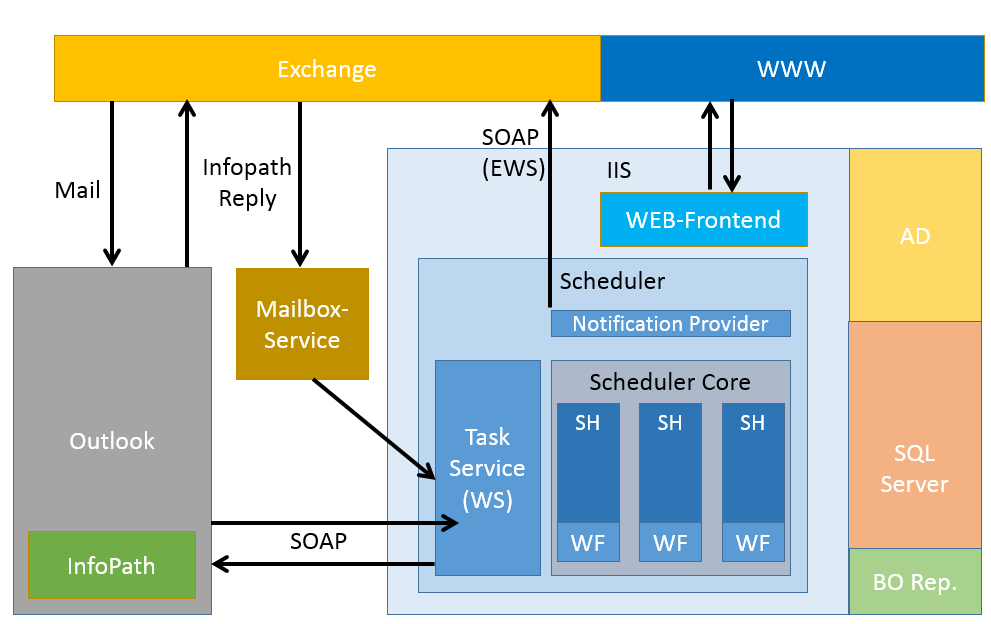}
	\caption{This Figure provides an overview of the whole architecture.}
	\label{fig:architecture}
\end{figure}

As discussed in the beginning, our objective was to build a prototype based on off-the-shelf infrastructure. Our infrastructure (see Fig.~\ref{fig:architecture}) consisted of two \emph{Microsoft Windows Server 2008 R2} machines. One of them was a domain controller, also hosting a DNS server, a \emph{Microsoft SQL Server 2008} instance which contained all databases for the prototype and an IIS web server, which hosted the \emph{Scheduler}. The other server provided an instance of \emph{Microsoft Exchange 2010}, as the notification of users was implemented by using the push mail functionality of \emph{Microsoft Exchange}. In our case, we implemented the Scheduler as a web application running on an IIS server, providing client interaction via web services using SOAP.

The web service makes the clients independent from our server core. This means that clients can be implemented using different technologies like .Net or Java for different platforms (Windows, Linux, Android etc.). We used \emph{Microsoft Outlook} and a web client (Browser) as an example for possible implementation scenarios.

At the very core of the server there are the workflows (labeled "WF" in Fig.~\ref{fig:architecture}). Each workflow is hosted by a \emph{SubjectHost} ("SH" in Fig.~\ref{fig:architecture}) which is a C\# class and takes care of functions like instantiating the workflow. All subject hosts together are hosted by the \emph{Scheduler Core}, which is also a C\# class. The \emph{Scheduler Core} coordinates the different subject hosts.

The \emph{Scheduler} holds all necessary functionality for the execution of S-BPM processes. Each process instance exists of several subject instances which are represented by the \emph{Subject Hosts} ("SH" in Fig.~\ref{fig:architecture}). Each \emph{Subject Host} hosts one workflow instance ("WF" in Fig.~\ref{fig:architecture}). The \emph{Subject Host} represents a logical and standardized environment for the execution of WF workflows.

This architecture implements the possibility of communication between individual activities within a workflow and the \emph{Scheduler}. Furthermore, the \emph{Subject Host} provides all essential functionalities to initialize, store and terminate workflows in a proper way. The \emph{Scheduler} manages all active and persisted processes, controls the communication between subjects and copes with \emph{Tasks} and \emph{Task} answers. For example, if one subject instance sends a message to another subject instance, the \emph{Scheduler} initializes a new \emph{Subject Host} (if not existing) and the \emph{Subject Host} creates either a new workflow instance or restores an existing instance from the database.

The components \emph{Task Service} and \emph{Notification Provider} implement the communication layer; the \emph{Task Service} is a C\# web service, which provides the necessary interfaces for client communication which is based on SOAP messages.

Methods provided by the \emph{Task Service} are:
\begin{itemize}
	\item List all processes for a specific user
	\item Start a new process
	\item List all open tasks for a specific user
	\item Receive a specific task
	\item Submit a specific task answer
\end{itemize}

In addition to the \emph{Task Service} for client communication, the \emph{Notification Provider} handles events of the \emph{Scheduler} directly. Such events are, for instance, triggered when an activity within a subject instance creates a \emph{Task} or a message. This event mechanism allows the implementation of a push mechanism. in our case we notify \emph{Microsoft Outlook} clients via \emph{Microsoft Exchange's} push-mail mechanism by sending E-Mails.

These three components together build the \emph{Scheduler}, the core of our application. The scheduler is hosted as a \emph{Microsoft IIS} web service, which means that the Internet Information Server starts the scheduler and keeps it running. The IIS also hosts a web front-end which can be used to upload processes and also provides a web client.

In addition to the IIS, we need an \emph{Active Directory}, a \emph{SQL Server} and a \emph{Business Object Repository}. The project was designed for \emph{Active Directory} integration so that no extra user database has to be maintained. Processes can be directly assigned to already defined organizational units, as defined in the \emph{Active Directory} and therefore to roles and users.

The \emph{SQL Server} is a crucial component as all persistent data is stored within a SQL database, including processes, workflows and workflow instances. While the IIS and the \emph{Active Directory} only work in conjunction with their corresponding other Microsoft components; the SQL server does not need to be a \emph{Microsoft SQL Server}.

Since we used \emph{Microsoft Outlook} as client application, user interaction happens via E-Mail messages. We implemented business objects using \emph{Microsoft InfoPath}, a form design software. These forms are attached to the E-Mails; They are stored in the \emph{Business Object Repository}, another component of our architecture. The client application communicates with the scheduler via exchange of SOAP messages through the \emph{Task Service}.

The last component shown in Fig.~\ref{fig:architecture} is the \emph{Mailbox Service}, which tracks the InfoPath forms, parses them and passes the information on to the scheduler.

\section{S-BPM Use-Case}

\subsection{General}

The use case selected to further explain the concept features a simple "internal order" process which is embedded into the organization using Microsoft \emph{Active Directory}. The process consists of two S-BPM subjects, an employee and a supervisor, who interact with each other.

\begin{figure}
	\centering
  	\includegraphics[scale=0.05]{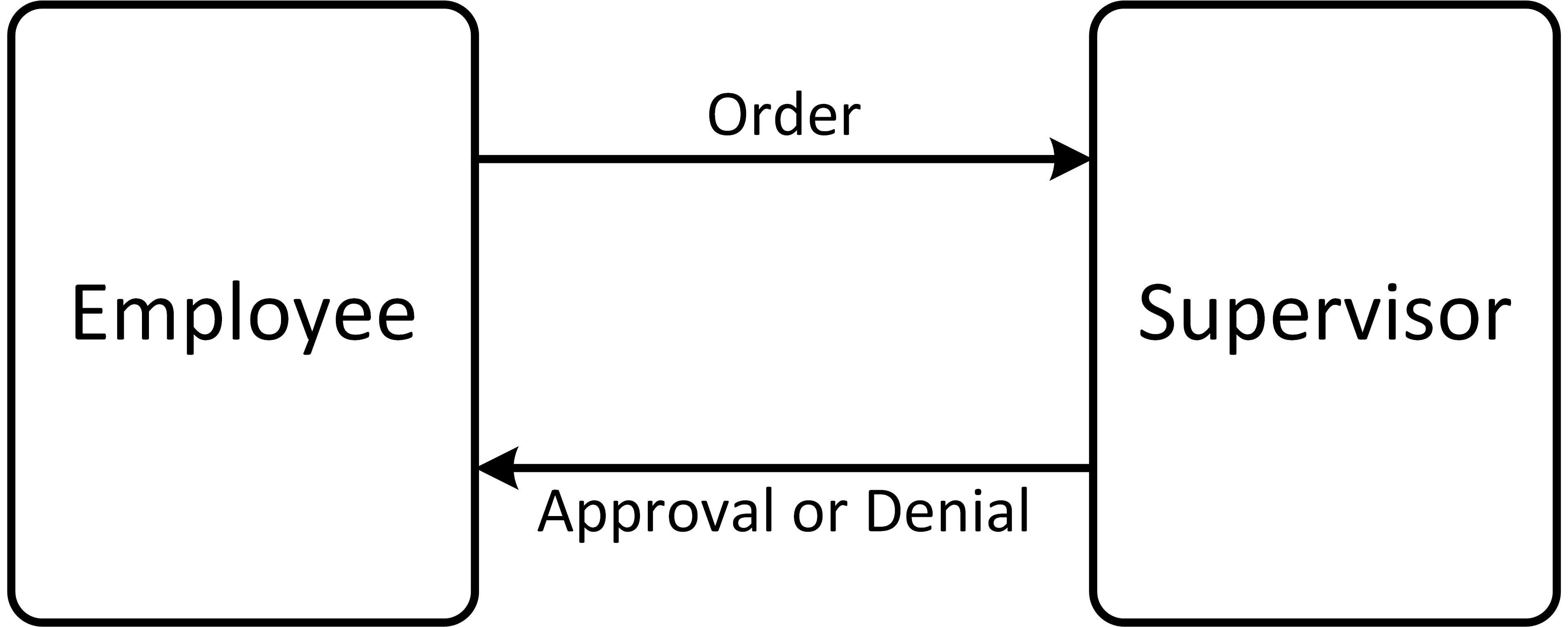}
	\caption{\emph{Subject Interaction Diagram} of the use-case example.}
	\label{fig:sid_io}
\end{figure}

As seen in the \emph{Subject Interaction Diagram} shown in Fig.~\ref{fig:sid_io}, the employee sends a message titled "Order" to the supervisor and receives an "Approval" or a "Denial" message in return. The use-case is explicitly simple for didactical reasons; nevertheless, it allows the demonstration of the full range of capabilities.

\begin{figure}
	\centering
  	\includegraphics[scale=0.05]{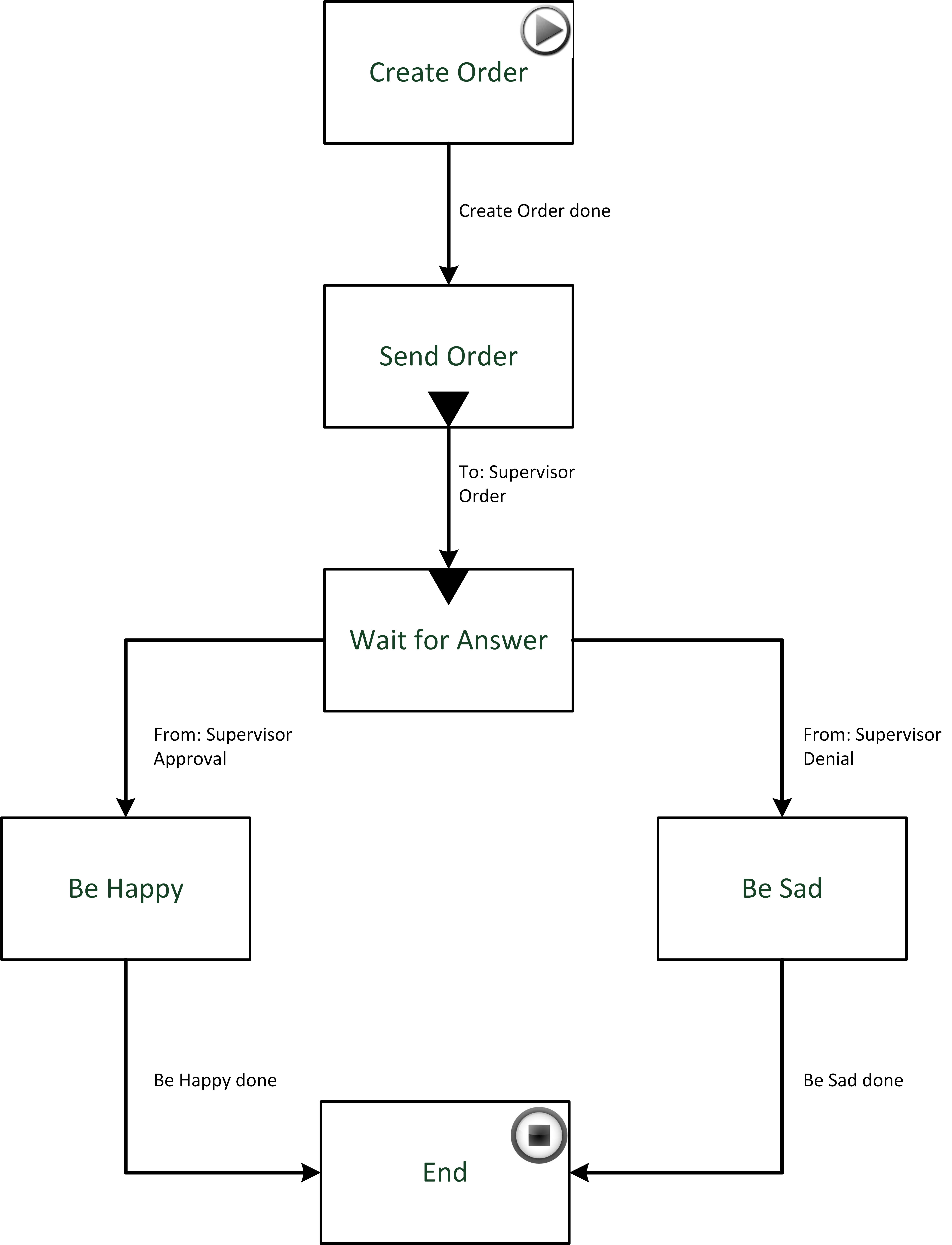}
	\caption{\emph{Subject Behavior Diagram} for the \emph{Employee}.}
	\label{fig:use_case_int_emp}
\end{figure}

\begin{figure}
	\centering
  	\includegraphics[scale=0.05]{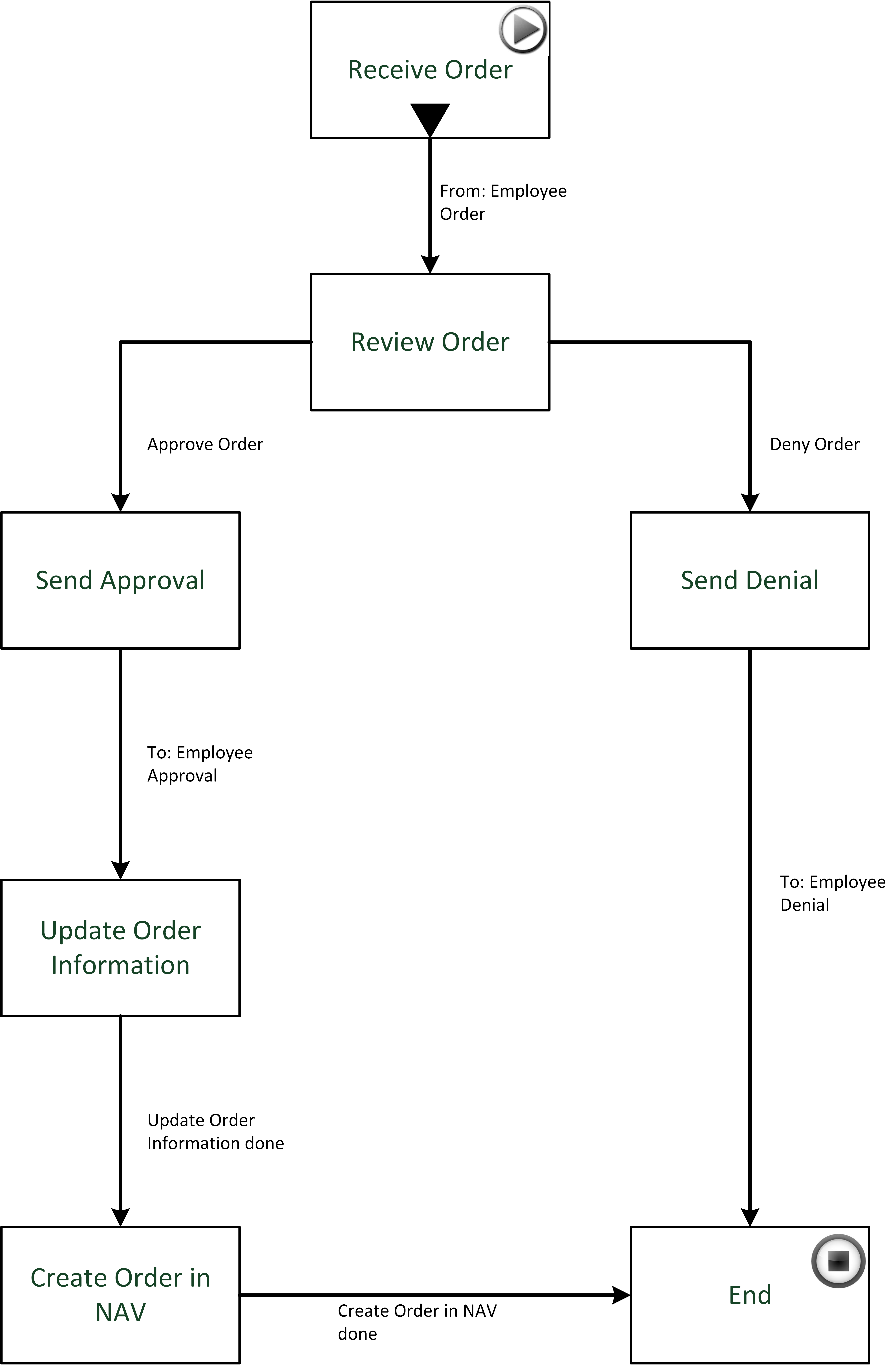}
\caption{The internal behavior of the Supervisor subject reflects the Employee subject, reacting to the received message and delivering one in return.}
\label{fig:use_case_int_sup}
\end{figure}

\begin{figure}
	\centering
  	\includegraphics[scale=0.05]{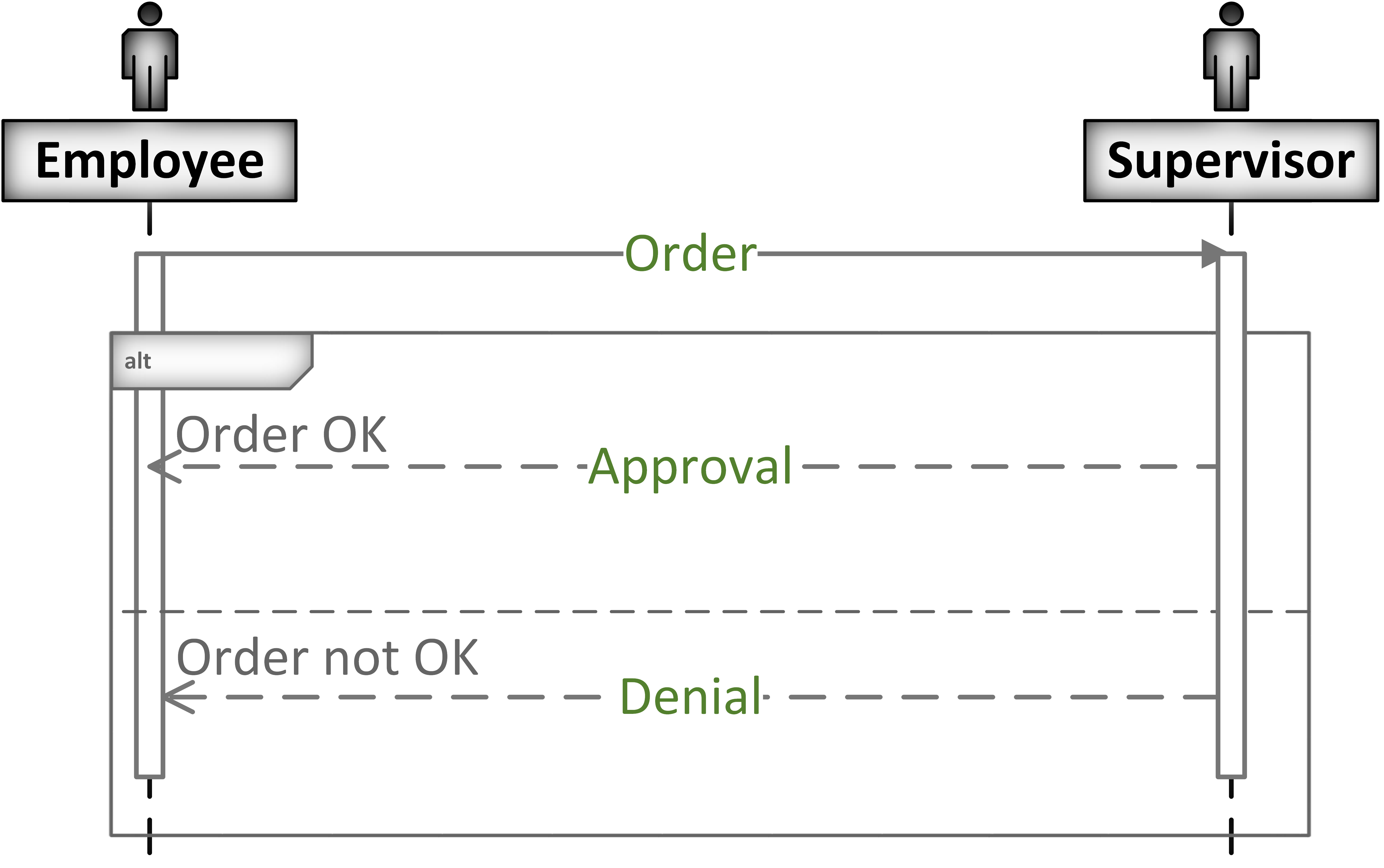}
	\caption{The UML \emph{User Interaction Diagram} of the use-case example.}
	\label{fig:use_case_interaction_v}
\end{figure}

The internal behavior of the \emph{Employee} subject, as described in Fig.~\ref{fig:use_case_int_emp}, is also simple: after creating the order, it is sent to the supervisor and the employee waits for an answer. The arriving answer is either an approval, in which case the employee is happy, or a denial, in which case the employee gets sad. Afterwards, the process ends for the employee. The internal behavior for the supervisor complements the behavior of the employee as shown in Fig.~\ref{fig:use_case_int_sup}: the supervisor receives the order and, after reviewing it, decides whether to approve or deny it. The result is then sent to the employee. In case of an approval, the supervisor must update the order information and enter it into the ERP system, in our case \emph{Microsoft Dynamics NAV}. Either way, the process also ends for the supervisor. Fig.~\ref{fig:use_case_interaction_v} shows an UML \emph{User Interaction Diagram} of the two subjects.

\begin{figure}
	\centering
  	\includegraphics[scale=0.1]{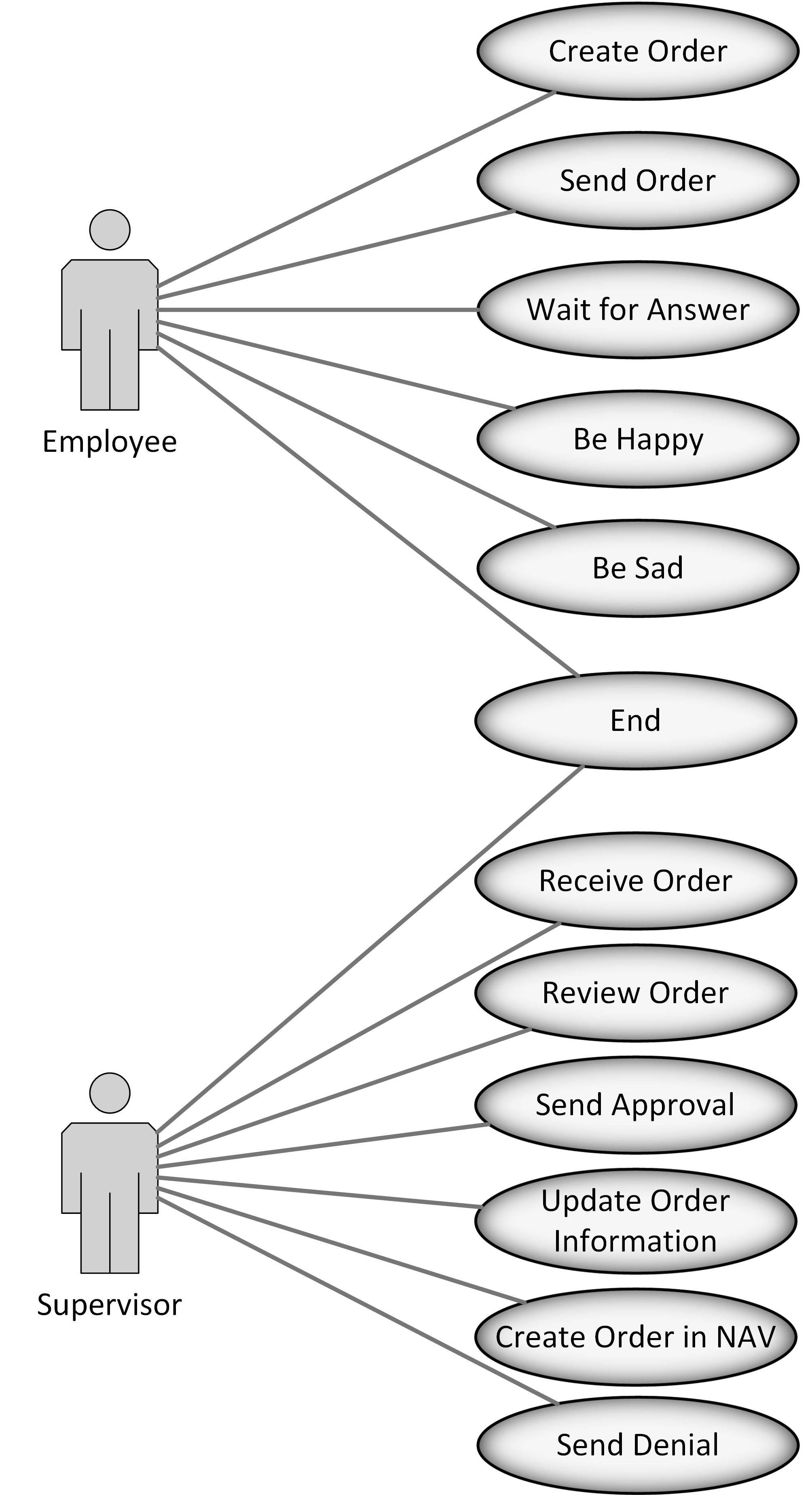}
	\caption{The use case as UML \emph{Use Case Diagram}.}
	\label{fig:umlusecase}
\end{figure}

\begin{figure}
	\centering
  	\includegraphics[scale=0.5]{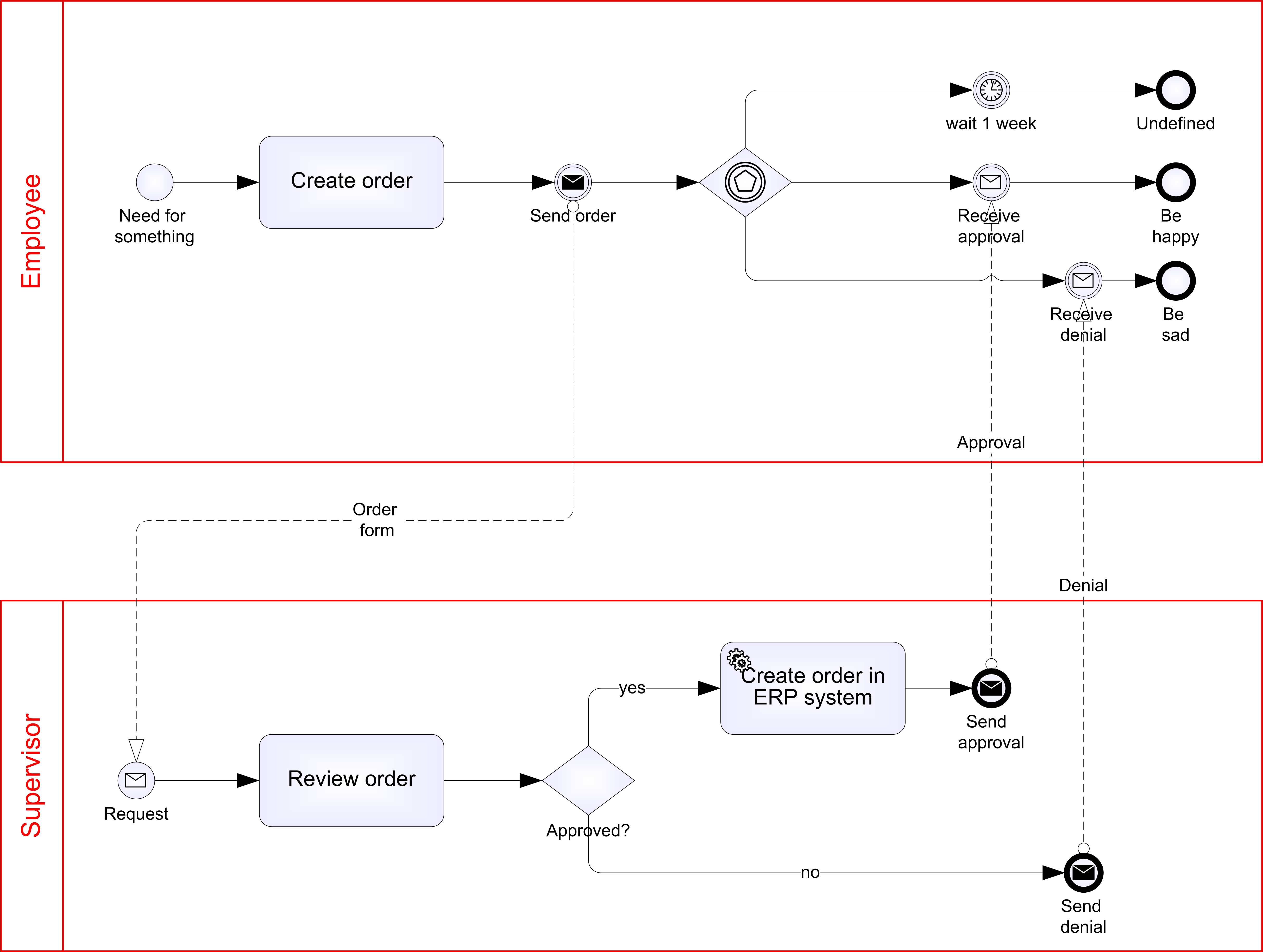}
	\caption{The basic behavior of the subjects of our sample process can be principally modeled as BPMN 2.0 processes. Be aware, that there are semantic and conceptional differences, as discussed in this work. This demonstration process also include a time event, which of course is also available in the S-BPM methodology.}
	\label{fig:bpmn}
\end{figure}

The use-case is also depicted as BPMN 2.0 process in Fig.~\ref{fig:bpmn}. A discussion about similarities and differences between S-BPM and BPMN models can be found in~\cite{Sneed.2012}; nevertheless, S-BPM supports process execution in a much more coherent way.

\begin{figure}
	\centering
  	\includegraphics[scale=0.6]{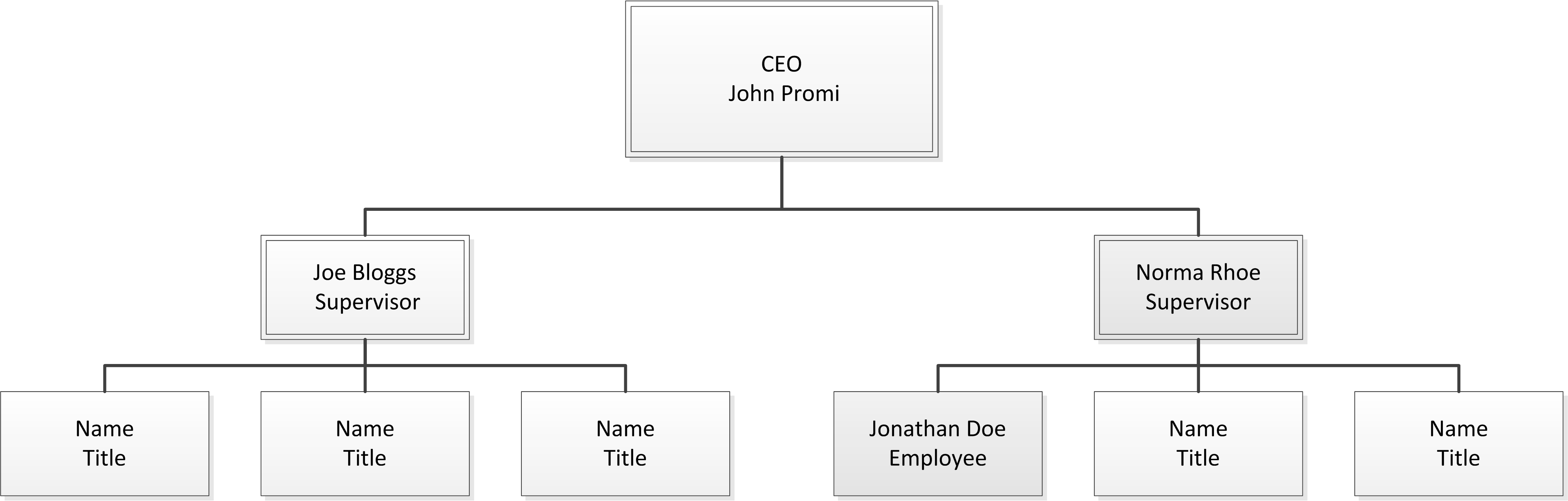}
	\caption{The fictional organization chart of the company PROMI.}
	\label{fig:promiorg}
\end{figure}

The fictional company in this use case is called PROMI. The organizational chart is shown in Fig.~\ref{fig:promiorg}. This organization chart is also mapped into the company's \emph{Active Directory} with the use of \emph{Active Directory Organizational Units}.

\subsection{Model Transformation and Upload}

\begin{figure}
	\centering
  	\includegraphics[scale=0.4]{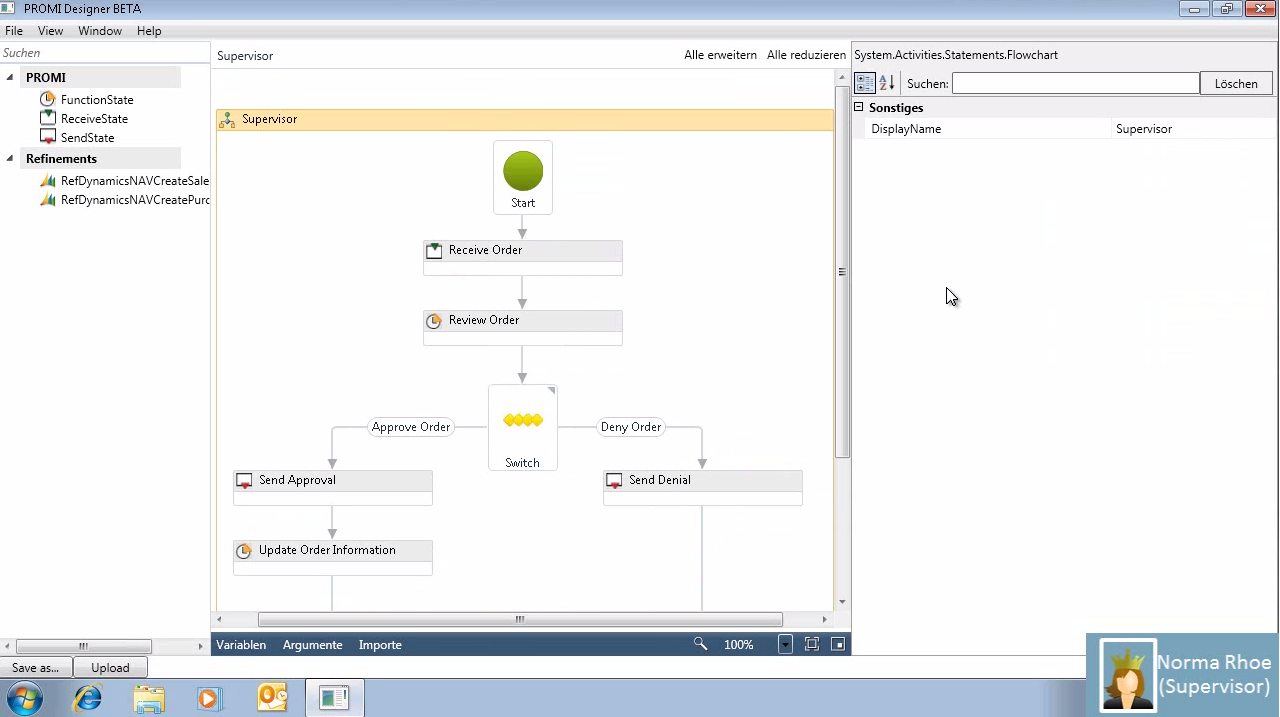}
	\caption{The \emph{PROMI Designer} allows modeling S-BPM processes as \emph{Windows Workflows}. This figure shows a part of the imported \emph{Supervisor} subject behavior}
	\label{fig:promidesigner}
\end{figure}

To enact a process it has to be uploaded into the execution platform; in our scenario we use the process suite from Metasonic (to prove compatibility of the interfaces) to define a S-BPM model. Behind the scene the Metasonic XML process model is translated into a XAML data structure, as described in the previous chapters. The upload is done via a web-frontend (see Fig.~\ref{fig:architecture}). Each subject (including its internal behavior) is automatically converted into a \emph{Windows Workflow}. Additionally, the workflows are then linked to their corresponding \emph{Organizational Units} in the \emph{Active Directory}, which ensures a role based access control. The uploaded and translated processes (now WF workflows) can be modified within the prototype platform via the usual development interfaces, as shown in Fig.~\ref{fig:promidesigner}.

\subsection{Execute the Process}

\subsubsection{Employee Subject - Request}

After uploading, the process is ready for use. In our case, Jonathan Doe (JD) needs a new notebook; so he has to use the new \emph{Internal Order} process to order a new one from his supervisor, Norma Rhoe  (NR). Depending on their role in an organization (realized via an active directory implementation) users can choose a process to start from a list; this is realized via an interface enhancement in \emph{Microsoft Outlook}.

In most cases a form has to be filled out (in our prototype we use \emph{Microsoft Infopath} for this purpose). In our case we have to collect some information for the order process (e.g. product to order). The next state of the process is to send the message to the supervisor subject (represented by NR). This is done using the e-mail infrastructure; technically, the form is sent as e-mail to the scheduler, which does all the work discussed previously: find the receiving subject (in our case NR) and persist the workflow until needed again.

\subsubsection{Supervisor Subject - Decision}

Any task for any process instance is sent to the inbox of the responsible subject(s) -- specially flagged, so that process e-mails (the messages) can be distinguished from "regular" mails. So, we use the same application (i.e. \emph{Microsoft Outlook} in our case) to process unstructured mails and structured mails, which are results of defined business processes.

At some point in time, the supervisor (NR) decides to process the request from her employee (JD). Offered tasks -- in the form of messages -- must be explicitly accepted from the open task list. Any informations are offered in a form and fields are readable only or writable. For example, a simple decision could be offered in form of radio buttons to reflect a yes or no decision. In our scenario the supervisor decides to grant the requested new laptop; the next state transition results in sending the requester (JD) a message of acceptance. Using the possibility of so called \emph{Refinements}\footnote{A \emph{Refinement} adds functionality to a \emph{Function State}, e.g. adding an additional class function using an appropriate programming language; in our case all code is written in C\#.} we included a simple web-service call to create a new order in an ERP system, \emph{Microsoft Dynamics NAV} in our case. The process is then ended for the supervisor subject (NR).

\subsubsection{Employee Subject - Answer}

The answer again is forwarded using the e-mail infrastructure; we also implemented a simple web frontend, so that any process step can alternatively be done using any web browser (e.g. using a smartphone). After accepting the answer from his supervisor (NR) on his request, the process also ends for the requester (JD). After both workflows are in their end state, the scheduler terminates both of them and the use case is finished.

\section{Summary and Conclusion}

The conclusions from this work are manifold:

Firstly, one main conclusion is, that the concept of S-BPM to understand all business processes as realizations of multi-agent systems directly leads to easy to understand process models which directly can be executed\footnote{In other words: "coordination of work is done through the exchange of messages".}.

Secondly, using standard off-the-shelf infrastructure solutions, a modeling and execution platform can be realized with well calculated effort; especially the concept and support of Microsoft's workflow technology offers a broad range of build in functionality for typical topics needed for long running business processes.

Thirdly, S-BPM offers a holistic approach to understand and study business processes; as Weske~\cite{Weske.2012} mentions, there are mainly three parties interested in the topic of business process management: business administration, communities in computer science, and software communities. We think, that S-BPM offers a common understanding and a research framework for all these parties; a coherent method to organize work with a human centric approach based on communication, a strong mathematical foundation, and a strong foundation for robust and scalable software implementations for distributed business processes (i.e. process choreographies) without semantical weaknesses.

In short, our S-BPM prototype implementation supplements unstructured with structured communication; we understand business processes as a framework to pre-define communication paths. We could demonstrate, that we can use the same communication tools to integrate both types of communication. In our case we simply used a standard e-mail client as human interaction interface for process communication. Further on, we want to point out, that our implementation of the S-BPM methodology also simply integrates any device into process communication (see for example \cite{Mueller.2012}), as long as it is able to communicate.

Currently the prototype implementation is transferred into a fully cloud based and service oriented infrastructure. We will further investigate, if functionality of the \emph{Scheduler} could be reduced or even omitted transferring more functionality from the central component to the distributed components (the subjects/agents).

\bibliography{Rass-Kotremba-Singer}

\begin{thebibliography}{10}
\providecommand{\url}[1]{\texttt{#1}}
\providecommand{\urlprefix}{URL }

\bibitem{Borger:2003}
B\"orger, E., St\"ark, R.F.: {Abstract State Machines. A Method for High-Level
  System Design and Analysis}. Springer (2003)

\bibitem{Borger:2011ib}
B{\"o}rger, E.: {A}pproaches to modeling business processes: a critical
  analysis of {B}{P}{M}{N}, workflow patterns and {Y}{A}{W}{L}. Software {\&}
  Systems Modeling  (2011)

\bibitem{Borgert.2011}
Borgert, S., Steinmetz, J., M\"uhlh\"auser, M.: {ePASS-IoS 1.1: Enabling
  Inter-enterprise Business Process Modeling by S-BPM and the Internet of
  Service Concept}. In: Schmidt, W. (ed.) S-BPM ONE -- Learning by Doing -
  Doing by Learning. CCIS, vol. 213, pp. 190--211. Springer (2011)

\bibitem{Chappell.2009}
Chappell, D.: {The Workflow Way - Understanding Windows Workflow Foundation}.
  http://www.davidchappell.com (2009)

\bibitem{Feldbacher:2011kf}
Feldbacher, P., Suppan, P., Schweiger, C., Singer, R.: {B}usiness {P}rocess
  {M}anagement: {A} {S}urvey among {S}mall and {M}edium {S}ized {E}nterprises.
  In: Schmidt, W. (ed.) S-BPM ONE -- Learning by Doing - Doing by Learning.
  CCIS, vol. 213, pp. 296--312. Springer Berlin Heidelberg (2011)

\bibitem{Fleischmann.1994}
Fleischmann, A.: {Distributed Systems: Software design and implementation}.
  Springer (1994)

\bibitem{Fleischmann:2013}
Fleischmann, A., Ra\ss, S., Singer, R.: {S-BPM Illustrated}. Springer (2013)

\bibitem{Fleischmann:2012va}
Fleischmann, A., Schmidt, W., Stary, C., Obermeier, S., B{\"o}rger, E.:
  {Subject-Oriented Business Process Management}. Springer (2012)

\bibitem{Hover.2013}
H\"over, K.M., Borgert, S., M\"uhlh\"auser, M.: {A Domain Specific Language for
  Describing S-BPM Processes}. In: Fischer, H., Schneeberger, J. (eds.) S-BPM
  ONE - Running Processes, Communications in Computer and Information Science,
  vol. 360, pp. 72--90. Springer Berlin Heidelberg (2013)

\bibitem{Milner.1980}
Milner, R.: {A Calculus of Communicating Systems}, LNCS, vol.~92. Springer
  (1980)

\bibitem{Mueller.2012}
M\"uller, H.: {Using S-BPM for PLC Code Generation and Extension of
  Subject-Oriented Methodology to All Layers of Modern Control Systems}. In:
  Stary, C. (ed.) S-BPM ONE - Scientific Research, Lecture Notes in Business
  Information Processing, vol. 104, pp. 182--204. Springer Berlin Heidelberg
  (2012)

\bibitem{Olbrich:2011zi}
Olbrich, T.J.: {W}hy {W}e {N}eed to {R}e-think {C}urrent {B}{P}{M} {R}esearch
  {I}ssues. In: Fleischmann, A., Schmidt, W., Singer, R., Seese, D. (eds.)
  {S}ubject-{O}riented {B}usiness {P}rocess {M}anagement. CCIS, vol. 138, pp.
  209--215. Springer (2011)

\bibitem{Puhlmann:2006mp}
Puhlmann, F.: {W}hy do we actually need the {P}i-{C}alculus for {B}usiness
  {P}rocess {M}anagement? LNI (85),  77--89 (2006)

\bibitem{Puhlmann:2005lf}
Puhlmann, F., Weske, M.: {U}sing the {P}i-{C}alculus for {F}ormalizing
  {W}orkflow {P}atterns. LNCS  3649,  153--168 (2005)

\bibitem{Silver.2011}
Silver, B.: {BPMN Method and Style}. Cody-Cassidy Press, 2 edn. (2011)

\bibitem{Silver.2012a}
Silver, B.: {Executable BPMN 2.0} (2012),
  \url{http://brsilver.com/executable-bpmn-2-0/}

\bibitem{Smith:2003zs}
Smith, H.: {B}usiness process management---the third wave: business process
  modelling language (bpml) and its pi-calculus foundations. Information and
  Software Technology  45(15),  1065--1069 (2003)

\bibitem{Smith:2004fy}
Smith, H., Fingar, P.: {W}orkflow is just a {P}i process (2004),
  \url{\url{www.bptrends.com}}

\bibitem{Smith.2007}
Smith, H., Fingar, P.: {Business Process Management: The Third Wave}.
  Meghan-Kiffer Press (2007)

\bibitem{Sneed.2012}
Sneed, S.H.: {Mapping Possibilities of S-BPM and BPMN 2.0}. In: Oppl, S.,
  Fleischmann, A. (eds.) Subject-Oriented Business Process Management. CCIS,
  vol. 284, pp. 91--105. Springer (2012)

\bibitem{Weske.2012}
Weske, M.: {Business Process Management: Concepts, Languages, Architectures}.
  Springer Berlin Heidelberg, 2 edn. (2012)

\bibitem{BPTrends.2012}
Wolf, C., Harmon, P.: {The State of Business Process Management 2012}. Report,
  BP Trends (2012)

\end{thebibliography}
\bibliographystyle{splncs03}

\end{document}